\documentclass[aps,pra]{revtex4}   
\newcommand {\be}{\begin{equation}}
\newcommand {\ee}{\end{equation}}
\newcommand {\bey}{\begin{eqnarray}}
\newcommand {\eey}{\end{eqnarray}}
\usepackage{epsfig}
\usepackage{amsfonts}
\begin{document}
\title{Exact BCS stochastic schemes for a time dependent many-body fermionic system}
\author{A. Montina}
\affiliation{Dipartimento di Fisica, Universit\`a di Firenze,
Via Sansone 1, 50019 Sesto Fiorentino (FI), Italy}
\author{Yvan Castin}
\affiliation{Laboratoire Kastler Brossel, \'Ecole Normale
Sup\'erieure, 24 rue Lhomond, 75005 Paris, France}

\begin{abstract}
The exact quantum state evolution of a fermionic gas with binary interactions is obtained as
the stochastic average of BCS-state trajectories. We find the most general Ito
stochastic equations which reproduce exactly the dynamics of
the system and we obtain some conditions to minimize the stochastic spreading of
the trajectories in the Hilbert space. The relation between the optimized equations 
and mean-field equations is analyzed. The method is applied to a simple two-site model. 
The simulations display effects that cannot be obtained in the mean-field 
approximation.
\end{abstract}
\maketitle

\section{Introduction}
The numerical solution of a time dependent many-body problem is a formidable task for large quantum systems. 
For example, for a system of an arbitrary number of fermions with $M$ possible modes
of the quantum field, e.g.\ in a lattice model, the dimension of the Hilbert space is $2^M$
so that both the computer time and the memory requirements scale exponentially with the
number of modes and becomes rapidly intractable when $M$ increases.
A similar situation occurs also in classical
mechanics, when we describe the dynamics as the evolution of a probability 
distribution. 

To circumvent the problem on the memory requirement in classical physics,
an approach is not to solve numerically the equation of motion
of the probability distribution, but to solve the statistical evolution of the variables
and to evaluate the mean value of some quantities over a finite number of realizations.
Such a Monte Carlo approach can be used also in quantum mechanics, the most famous example
being the Path Integral Monte Carlo based on the Feynman's path integral
formulation \cite{pimc}. 
However, apart from notable exceptions (such as the imaginary time evolution of 
bosons with real Hamiltonians, or the imaginary time evolution of some models of fermions \cite{determinantal_MC}),
for a general quantum problem, the known Quantum Monte Carlo methods do not solve the computer time
problem, which remains exponentially long because an exponentially large number
of Monte Carlo realizations is usually required \cite{Troyer}.
For fermions, this problem is the celebrated sign problem, which has been the subject
of many efforts, both for real-time and 
imaginary-time simulations~\cite{sign0,sign1,plimak,sign2,sign3,sign4,sign5,sign6,sign7,sign8}.

In Monte Carlo techniques with path integral, the randomly generated states are generally 
mutually orthogonal. In the original formulation of Feynman, they are the eigenvectors of the 
particle coordinates. As an alternative, we can use to randomly explore an over-complete set of states.
Since the dynamics of degenerate bosonic gas with weak interactions is approximatively 
described by the evolution of a Hartree-Fock state, it can be convenient to evaluate
the exact dynamics using a superposition of paths of Hartree-Fock states~\cite{cdc}. 
A similar approach has been used for fermion systems~\cite{juillet}. In 
Reference~\cite{cdc} the case of bosonic coherent states is also studied.

The number of random paths necessary to describe the dynamics can be reduced
by increasing the number of elements of the over-complete set. In the extreme limit
where every state of the Hilbert state is an element of the explored set, the dynamics
can be described with a single, deterministic path: this corresponds to solving directly the Schr\"odinger 
equation, but this faces again the memory problem. As an intermediate possibility, we can choose a set of
elements whose single path is a better approximation to the exact solution than the 
Hartree-Fock ansatz, but which is still numerically tractable. Attractive interactions
in a fermionic gas can lead to the condensation of Cooper pairs in the superfluid state,
as currently investigated experimentally in atomic gases close to a Feshbach
resonance \cite{experiments}. It is expected that such a superfluid state is
reasonably well described by a BCS-state, much better indeed than by a Hartree-Fock state.
For this reason we study here an exact stochastic approach with BCS states.

In this article we consider the dynamics in real-time of a system of fermions with 
binary interactions on a spatial lattice. The Hamiltonian is
\be\label{hami}
\hat {\cal H}=\sum_{kl}h_{kl}\hat c_k^\dagger\hat c_l+
\frac{1}{2}\sum_{kl}V_{kl}\hat c_k^\dagger\hat c_l^\dagger\hat c_l
\hat c_k,
\ee
$h$ and $V$ are hermitian and real symmetric matrices, respectively, and $\hat c_k$,
$\hat c_k^\dagger$ are Fermi annihilation and creation operators. 
The mode index $k$ labels the spin state $\sigma_k$ and the lattice node in position $r_k$.
In what follows, we shall denote as $m_s$ the total number of lattice nodes.

We wish to obtain the dynamical evolution of the quantum state 
as the average of stochastic trajectories of BCS states. 
The exact evolution is achieved by averaging an infinite number of
stochastic trajectories. 
The BCS state ansatz that we use is
\be
\label{ansatz}
|\Omega,\gamma\rangle\equiv \Omega S(\gamma)|0\rangle\equiv
\Omega \, e^{\frac{1}{2}\sum_{kl}\gamma_{kl}\hat c_k^\dagger\hat c_l^\dagger}
|0\rangle,
\ee 
$\gamma$ being an antisymmetric matrix, involving for spin $1/2$ fermions
a number of variables $2m_s(2m_s-1)/2$,
and $\Omega$ being a multiplicative complex variable.
Note that the state in Eq.~(\ref{ansatz}) is in general not normalized. 
We shall consider the case where both $\gamma$ and $\Omega$ are stochastic
variables solving Ito stochastic equations~\cite{why_not_bad}.

In Section~\ref{sect3} we find the necessary and sufficient conditions on 
the stochastic equations in order to have an exact description of 
the dynamics.
These constraints do not fix univocally the stochastic scheme, thus we shall use this 
freedom to reduce the statistical spreading of the trajectories. In Section~\ref{sec_expl}
we construct explicit stochastic schemes.
The growth rate of the spreading is evaluated and an upper limit for the 
statistical error on the observables is established, which shows that
the statistical uncertainty is finite at every finite time. 
In Section~\ref{sect4} the stochastic approach is illustrated on a two-site model.

\section{Stochastic equations}
\label{sect3}

\subsection{Conditions for the stochastic evolution to be exact}
We want to evaluate exactly the quantum state evolution using a superposition
of the BCS states $|\Omega,\gamma\rangle\equiv\Omega|\gamma\rangle$
with a stochastic evolution of $\gamma$ and $\Omega$.
For an infinitesimal variation of $\gamma$ and $\Omega$ we calculate
the variation of the ansatz by expanding $|\Omega+\Delta\Omega,\gamma
+\Delta\gamma\rangle$ in powers of $\Delta\Omega$ and $\Delta\gamma$:
we have from Eq.~(\ref{pro1}) that
\be\label{DAnsatz}
\Delta|\Omega,\gamma\rangle=
\left[\frac{\Delta\Omega}{\Omega}+
\frac{1}{2}\sum_{ij}\Delta\gamma_{ij}\hat c_i^\dagger
\hat c_j^\dagger+\frac{1}{8}\sum_{ijkl}\Delta\gamma_{ij}\Delta\gamma_{kl}
\hat c_i^\dagger\hat c_j^\dagger\hat c_k^\dagger\hat c_l^\dagger+
\frac{1}{2\Omega}\sum_{ij}\Delta\Omega\Delta\gamma_{ij}\hat c_i^\dagger
\hat c_j^\dagger+...
\right]|\Omega,\gamma\rangle,
\ee
On the other side, the Hamiltonian evolution during $\Delta t$ 
of the state equal to $|\Omega,\gamma\rangle$ at time $t$ is given
to first order in $\Delta t$ by Schr\"odinger's equation:
\be
\label{schrod}
-\mathrm{i} \hat{\cal H} \Delta t |\Omega,\gamma\rangle=
\left[\frac{\mathrm i}{2}\sum_{ijkl}V_{ij}\gamma_{ik}\gamma_{jl}\hat c_i^\dagger\hat c_j^\dagger
\hat c_k^\dagger\hat c_l^\dagger\Delta t
-{\mathrm i}\sum_{ij}\left(\frac{1}{2}V_{ij}\gamma_{ij}+\sum_kh_{ik}\gamma_{kj}\right)
\hat c_i^\dagger\hat c_j^\dagger\Delta t\right]|\Omega,\gamma\rangle
\ee
where we used Eq.~(\ref{pro2}) to express $\hat{c}|\Omega,\gamma\rangle$
in terms of $\hat{c}^\dagger|\Omega,\gamma\rangle$ and where we took $\hbar=1$.
If $\gamma$ and $\Omega$ satisfy a deterministic equation, it is obvious that the
first term of the right hand side of Eq.~(\ref{schrod}) does not in general coincide
with the third term of the right hand side of 
Eq.~(\ref{DAnsatz}).
As we shall prove, Eq.(\ref{schrod}) and Eq.(\ref{DAnsatz}) can become equal when we consider
stochastic equations and we average Eq.(\ref{DAnsatz}) over every possible realization of
the stochastic variation during $\Delta t$:
\be
\overline{\Delta |\Omega,\gamma\rangle} = 
-\mathrm{i} \hat{\cal H} \Delta t |\Omega,\gamma\rangle.
\ee
Since $S(\gamma)$ is invertible, we have to find
a stochastic equation for $\Omega$ and $\gamma$ that satisfies the equality
\bey\nonumber
\left[\frac{\overline{\Delta\Omega}}{\Omega}+
\frac{1}{2}\sum_{ij}\overline{\Delta\gamma_{ij}}\hat c_i^\dagger
\hat c_j^\dagger
+\frac{1}{8}\sum_{ijkl}\overline{\Delta\gamma_{ij}\Delta\gamma_{kl}}
\hat c_i^\dagger\hat c_j^\dagger\hat c_k^\dagger\hat c_l^\dagger+
\frac{1}{2\Omega}\sum_{ij}\overline{\Delta\Omega\Delta\gamma_{ij}}\hat c_i^\dagger
\hat c_j^\dagger\right]|0\rangle=   \\
\label{equality}
\left[\frac{\mathrm i}{2}\sum_{ijkl}V_{ij}\gamma_{ik}\gamma_{jl}\hat c_i^\dagger\hat c_j^\dagger
\hat c_k^\dagger\hat c_l^\dagger\Delta t
-\mathrm i\sum_{ij}\left(\frac{1}{2}V_{ij}\gamma_{ij}+\sum_kh_{ik}\gamma_{kj}\right)
\hat c_i^\dagger\hat c_j^\dagger\Delta t\right]|0\rangle.
\eey
Equation~(\ref{equality}) is equivalent to the following ones,
\bey\label{equality_1}
\frac{\overline{\Delta\Omega}}{\Omega}=0
\\
\label{equality_2}
\overline{\Delta\gamma_{ij}}=-\frac{1}{\Omega}\overline{\Delta\Omega\Delta\gamma_{ij}}
-{\mathrm i}V_{ij}\gamma_{ij}\Delta t-{\mathrm i}\sum_kh_{ik}\gamma_{kj}\Delta t+
{\mathrm i}\sum_kh_{jk}\gamma_{ki}\Delta t \\
\sum_{\textrm{permutation of }ijkl}(-1)^p\left[\frac{1}{8}
\overline{\Delta\gamma_{ij}\Delta\gamma_{kl}}
-\frac{\mathrm i}{2}V_{ij}\gamma_{ik}\gamma_{jl}\Delta t\right]=0
\eey
where $(-1)^p$ is the signature of the permutation.
The first equation implies that the deterministic term of $\Omega$ is zero. The second equation 
gives the deterministic term for $\gamma$, the last one gives a condition for the noise term of $\gamma$. 
This last condition can be written explicitly 
\bey
\nonumber
\overline{\Delta\gamma_{ij}\Delta\gamma_{kl}}+\overline{\Delta\gamma_{jk}\Delta\gamma_{il}}
+\overline{\Delta\gamma_{ik}\Delta\gamma_{lj}}+{\mathrm i}\Delta t
(V_{ij}+V_{kj}+V_{il}
+V_{kl})\gamma_{ik}\gamma_{lj}  \\
\label{equality_3}
+{\mathrm i}\Delta t (V_{ik}+V_{kj}+V_{il}+V_{jl})\gamma_{ij}\gamma_{kl}
+{\mathrm i}\Delta t (V_{ik}+V_{lk}+V_{ij}+V_{lj})\gamma_{jk}\gamma_{il}=0
\eey
Note that this equation is automatically fulfilled when two of the four indices $ijkl$
are equal.

\subsection{Growth of the statistical error}
To estimate the statistical error of the method,
we consider the growth rate of the mean squared distance between the true state of the system and a single
realization of the stochastic ansatz, $\overline{\Delta||\psi\rangle-|\Omega,\gamma\rangle|^2}=
\overline{\Delta M}$, where
\be
M = \langle\Omega,\gamma|\Omega,\gamma\rangle.
\ee
This will allow to prove that the statistical error remains finite at all finite evolution times and
this will provide a strategy to identify optimal stochastic schemes in trying to minimize the growth
rate $\overline{\Delta M} /M$.

To the first order in $\Delta t$,
\be\label{err_growth}
\overline{\Delta M}=
\overline{(\Delta\langle\Omega,\gamma|)}|\Omega,\gamma\rangle+
\langle\Omega,\gamma|\overline{(\Delta|\Omega,\gamma\rangle)}+
\overline{(\Delta\langle\Omega,\gamma|)(\Delta|\Omega,\gamma\rangle)}.
\ee
In the right hand side, the sum of the first two terms gives exactly zero, since
by construction $\overline{\Delta|\Omega,\gamma\rangle}= -{\mathrm i} \Delta t \hat{\cal H} |\Omega,\gamma\rangle/\hbar$.
In the last term, we can replace $\Delta |\Omega,\gamma\rangle$ by its stochastic component:
\bey\nonumber
\Delta|\Omega,\gamma\rangle^{\rm stoch}  &\equiv & \Delta|\Omega,\gamma\rangle- \overline{\Delta|\Omega,\gamma\rangle} \\
\label{stoch_comp}
&=& \left[\frac{\Delta \Omega}{\Omega} + \frac{1}{2} \sum_{ij} \Delta\gamma_{ij}^{\rm stoch}\,
\hat{c}^\dagger_i \hat{c}^\dagger_j \right] |\Omega,\gamma\rangle,
\eey
where we used the fact that $\Delta \Omega$ is purely stochastic and where
$\Delta\gamma_{ij}^{\rm stoch}$ is the stochastic part of $\Delta\gamma_{ij}$.
Equation~(\ref{err_growth}) leads to
\be
\frac{\overline{\Delta M}}{M} = \frac{||\Delta|\Omega,\gamma\rangle^{\rm stoch}||^2}{\langle \Omega,\gamma|\Omega,\gamma\rangle}
\label{growth}
\ee
which can be evaluated using Wick's theorem:
\be
\label{growth_util}
\frac{\overline{\Delta M}}{M} =\left|\frac{\Delta\Omega}{\Omega}+
\frac{1}{2}\sum_{ij} \Delta \gamma_{ij}^{\rm stoch}\langle\hat{c}_i^\dagger\hat{c}_j^\dagger\rangle\right|^2
+ \frac{1}{2}\sum_{ijkl} \overline{\Delta\gamma_{ij}^* \Delta\gamma_{kl}}
\langle \hat{c}_i \hat{c}_k^\dagger\rangle 
\langle \hat{c}_j \hat{c}_l^\dagger\rangle.
\ee
where the expectation value is taken in the ansatz, 
$\langle \ldots \rangle = \langle \Omega,\gamma| 
\ldots |\Omega,\gamma\rangle/M$.
A clear step for the minimization of the error growth is to choose $\Delta\Omega$
in order to set to zero the first term in the right hand side of the above expression:
\be
\label{clear_step}
\frac{\Delta\Omega}{\Omega} = - \frac{1}{2}\sum_{kl} \Delta \gamma_{kl}^{\rm stoch}
\langle\hat{c}_k^\dagger\hat{c}_l^\dagger\rangle.
\ee
We shall always choose $\Delta\Omega$ in this way in what follows.
It is then easy to show [see Eq.~(\ref{stoch_comp})] that
\be
\overline{\Delta M}=\Delta M,
\ee
i.e., the stochastic terms of $\Delta M$ are exactly zero.
Furthermore the deterministic part of $\Delta\gamma$ is now slaved to the stochastic part
of $\Delta\gamma$, according to Eq.(\ref{equality_2}): the only increments that remain to be
specified to fully determine the stochastic scheme are $\Delta\gamma_{ij}^{\rm stoch}$,
and this we shall do in the next section.

\section{Explicit exact stochastic schemes}
\label{sec_expl}

\subsection{Our solution for an arbitrary interaction potential}
\label{subsec:sol}

This most general solution relies on the following ansatz for the stochastic increment:
\begin{equation}
\label{ansate}
\Delta\gamma_{ij}^{\rm stoch} = (\Delta f_i+\Delta f_j) \gamma_{ij}
\end{equation}
where the noise terms $\Delta f_k$ are independent of $\gamma_{ij}$.
Inserting this ansatz in the validity condition Eq.(\ref{equality_3}), 
we find that if the noise terms have the following correlation function,
\begin{equation}
\overline{\Delta f_i \Delta f_j} = -{\mathrm i} V_{ij} \Delta t, 
\end{equation}
this validity condition is satisfied \cite{suff}.
Since the matrix $V_{ij}$ is real symmetric it can be diagonalized; a noise
having this correlation function may then be explicitly constructed using
the corresponding eigenbasis. 

In the specific case of a discrete $\delta$ interaction potential
between two opposite spin components:
\begin{equation}
V_{ij} = V_0 \delta_{r_i,r_j} \delta_{\sigma_i,-\sigma_j}
\label{delta}
\end{equation}
where $r_i$ and $\sigma_i$ are the lattice position and the spin component of the mode of index $i$, the
following explicit noise may be used:
\begin{equation}
f_i = (-\mathrm{i} V_0)^{1/2} \left[\Delta\xi_{r_i} \delta_{\sigma_i,\uparrow}
+\Delta\xi_{r_i}^* \delta_{\sigma_i,\downarrow}\right]
\end{equation}
where the $\Delta\xi_{r}$'s are statistically independent complex Gaussian noises of variance $\Delta t$.
For this specific noise implementation, $\overline{f^*_i f_j}=|V_0|\Delta t
\,\delta_{ij}$, so that the growth rate of 
the statistical error can be expressed by the simple formula \cite{aide}:
\begin{equation}
\label{crescita_generale}
\frac{\Delta M}{M}= |V_0| \Delta t \sum_k \langle \hat{c}_k \hat{c}_k^\dagger \rangle
\langle \hat{c}_k^\dagger \hat{c}_k \rangle \leq \frac{1}{2} |V_0| \Delta t\, m_s
\end{equation}
where $m_s$ is the number of lattice nodes and where we used the Eqs.~(\ref{cpc},\ref{matrid},
\ref{cc}).
$\Delta M/(M\Delta t)$ has a constant as an upper bound, thus the norm squared of $|\Omega,\gamma\rangle$ 
is bounded at every time by $\exp(|V_0|m_s t/2)$ times its initial value.
A similar bound was derived in the stochastic Hartree-Fock scheme in
\cite{juillet}, with a larger exponent \cite{how_large}.
Consequently, 
the Monte Carlo statistical variance of an observable $O$
is finite when ${\mathrm Tr}[O^2]$ is finite~[see Ref.~\cite{caru1}, Eq.~(30)].

To be complete we also give the corresponding deterministic part of the evolution of $\gamma$:
\be
\overline{\Delta\gamma_{ij}}/\Delta t=
-{\mathrm i}V_{ij}\gamma_{ij}
-{\mathrm i}\sum_k(h_{ik}\gamma_{kj}+h_{jk}\gamma_{ik})
-{\mathrm i}\sum_k(V_{ik}+V_{jk})\langle \hat c_k^\dagger\hat c_k\rangle\gamma_{ij}.
\ee
We note that the last sum over $k$ in the righthand side is simply
the Hartree mean-field term. 

\subsection{Case of an off-diagonal $\gamma$ and a delta interaction}

We now restrict to useful limiting cases where the one-body Hamiltonian $h$ is
spin-diagonal, the interaction potential is an on-site
discrete $\delta$ between two distinct spin components,
as defined in Eq.(\ref{delta}),
and where the matrix $\gamma$ in the ansatz has
initially zero matrix elements between identical spin components.
We have then identified exact stochastic schemes that preserve at any time
this block off-diagonal structure of $\gamma$.

\subsubsection{The solution that we have found with minimal error growth}
\label{subsubsec:optimal}

A general strategy to find the `best' stochastic scheme among a very large number
of possibilities is to try to minimize the growth rate of the statistical error
Eq.(\ref{growth}). Whereas this program is easily fulfilled for bosons
\cite{cdc} it seems to be more difficult to achieve for the stochastic
BCS ansatz. Here we report the solution that we have found with minimal error
growth. It is possible that a better solution exists. The stochastic increment is 
given by
\begin{equation}
\Delta \gamma_{\uparrow r_i, \downarrow r_j}^{\rm stoch} = ({\mathrm i} V_0)^{1/2}
(\Delta \xi_{r_i} -\Delta \xi_{r_j}) \gamma_{\uparrow r_i, \downarrow r_j}
\end{equation}
where the $\Delta\xi_r$ are $m_s$ independent real Gaussian noises of variance $\Delta t$.
The corresponding growth rate of the statistical error is exactly given by
\begin{equation}
\label{growth3}
\frac{\Delta M}{M}=
\Delta t |V_0|\, \sum_{r}\left[\langle\hat c^\dagger_{\uparrow r}\hat c_{\uparrow r}\rangle
\langle\hat c_{\uparrow r}\hat c^\dagger_{\uparrow r}\rangle+
\langle\hat c^\dagger_{\downarrow r}\hat c_{\downarrow r}\rangle
\langle\hat c_{\downarrow r}\hat c^\dagger_{\downarrow r}\rangle-
2\langle\hat c^\dagger_{\uparrow r}\hat c_{\downarrow r}^\dagger\rangle
\langle\hat c_{\downarrow r}\hat c_{\uparrow r}\rangle\right]
\end{equation}
which is indeed smaller than the general result Eq.(\ref{crescita_generale}) because
of the occurrence of a negative term involving anomalous averages.
In this scheme the deterministic part of the evolution of $\gamma$ is given by
\be
\overline{\Delta\gamma_{\uparrow r_i, \downarrow r_j}}/\Delta t=
-{\mathrm i}V_0\gamma_{\uparrow r_i,\downarrow r_j}[\langle \hat c_{\downarrow r_i}^\dagger\hat c_{\downarrow r_i}\rangle-
\langle \hat c_{\uparrow r_i}^\dagger\hat c_{\uparrow r_i}\rangle-\{i\leftrightarrow j\} ]
-{\mathrm i}V_0\delta_{r_i,r_j}\gamma_{\uparrow r_i,\downarrow r_j}
-{\mathrm i}\sum_{r_k}(h_{\uparrow r_i,\uparrow r_k}\gamma_{\uparrow r_k
,\downarrow r_j}+h_{\downarrow r_j,\downarrow r_k}\gamma_{\uparrow r_i,\downarrow r_k})
\ee

\subsubsection{The solution that we have found with minimal memory requirement}
\label{subsubsec:low_mem}

In the case when the one-body Hamiltonian $h$ is totally spin
independent and when 
the off-diagonal block $\gamma_{\uparrow r_i,\downarrow r_j}$
is initially a symmetric or antisymmetric matrix (under the exchange of $r_i$ and $r_j$), we have found
a stochastic scheme which preserves this symmetry property at all times, which allows
to save a factor of two on the memory requirement;
\be
\Delta \gamma_{\uparrow r_i, \downarrow r_j}^{\rm stoch} = i(i V_0)^{1/2}
(\Delta \xi_{r_i} +\Delta \xi_{r_j}) \gamma_{\uparrow r_i, \downarrow r_j} 
\ee
where the $\Delta\xi_r$ are $m_s$ independent real Gaussian noises of variance $\Delta t$.
In this case, the growth rate of the statistical error is
\begin{equation}
\label{growth4}
\frac{\Delta M}{M}=        
\Delta t |V_0|\, \sum_{r}\left[\langle\hat c^\dagger_{\uparrow r}\hat c_{\uparrow r}\rangle    
\langle\hat c_{\uparrow r}\hat c^\dagger_{\uparrow r}\rangle+  
\langle\hat c^\dagger_{\downarrow r}\hat c_{\downarrow r}\rangle  
\langle\hat c_{\downarrow r}\hat c^\dagger_{\downarrow r}\rangle+  
2\langle\hat c^\dagger_{\uparrow r}\hat c_{\downarrow r}^\dagger\rangle  
\langle\hat c_{\downarrow r}\hat c_{\uparrow r}\rangle\right]\leq \Delta t |V_0| m_s
\end{equation} 
which is larger than for the two previous schemes.
In this scheme the deterministic part of the evolution of $\gamma$ is given by
\be
\overline{\Delta\gamma_{\uparrow r_i, \downarrow r_j}}/\Delta t=
-{\mathrm i}V_0\gamma_{\uparrow r_i,\downarrow r_j}\left[\langle \hat c_{\downarrow r_i}^\dagger\hat c_{\downarrow r_i}\rangle+
\langle \hat c_{\uparrow r_i}^\dagger\hat c_{\uparrow r_i}\rangle+\{i\leftrightarrow j\} \right]
-{\mathrm i}V_0\delta_{r_i,r_j}\gamma_{\uparrow r_i,\downarrow r_j}
-{\mathrm i}\sum_{r_k}(h_{\uparrow r_i,\uparrow r_k}\gamma_{\uparrow r_k
,\downarrow r_j}+h_{\downarrow r_j,\downarrow r_k}\gamma_{\uparrow r_i,\downarrow r_k})
\ee

\subsection{Link with the mean-field approximation}

In the three explicit stochastic schemes given in this section, the deterministic part of
the evolution for $\gamma_{ij}$ did not coincide with the mean-field evolution.
This in contrast with the optimized stochastic Hartree-Fock schemes obtained for bosons
\cite{cdc} and for fermions \cite{juillet}.
For a general stochastic BCS ansatz with the optimizing choice Eq.(\ref{clear_step}),
we found the following relation between the deterministic evolution of
$\gamma_{ij}$ and its mean field evolution (given in Appendix~\ref{appen:mf}):
\be
\overline{\Delta\gamma_{ij}}-\Delta\gamma_{ij}^{\mbox{\scriptsize
mean field}} =
\sum_{kl} \overline{\Delta\gamma_{ik}\Delta\gamma_{jl}}
\langle \hat{c}_k^\dagger \hat{c}_l^\dagger\rangle
\ee
by inserting the expression Eq.~(\ref{clear_step}) of $\Delta\Omega$ into
Eq.~(\ref{equality_2}) and by using Eq.~(\ref{equality_3}) to eliminate
$\overline{\Delta\gamma_{ij}\Delta\gamma_{kl}}$.
This shows that finding a stochastic scheme where the deterministic and mean
field evolutions coincide is not straightforward.

\section{Stochastic approach for a two-site system}
\label{sect4}
In order to illustrate the method, we apply it
 to a simple system with two sites, corresponding to the Hamiltonian
\be
\hat{\cal H}=
\frac{1}{2}\sum_\sigma\left[\hat c_{\sigma 1}^\dagger\hat c_{\sigma 2}+\hat c_{\sigma 2}^\dagger\hat c_{\sigma 1}\right]
+V\left[
\hat c_{\uparrow 1}^\dagger\hat c_{\downarrow 1}^\dagger \hat c_{\downarrow 1}\hat c_{\uparrow 1}+
\hat c_{\uparrow 2}^\dagger\hat c_{\downarrow 2}^\dagger \hat c_{\downarrow 2}\hat c_{\uparrow 2}\right]
\ee
where the spin index $\sigma$ takes the values $\uparrow$ and $\downarrow$.
There is no interparticle interaction
when the two particles are in different wells.
A physical system that may be described by this model 
is a set of two Fermi particles in a double-well potential. 

\begin{figure}
\epsfig{figure=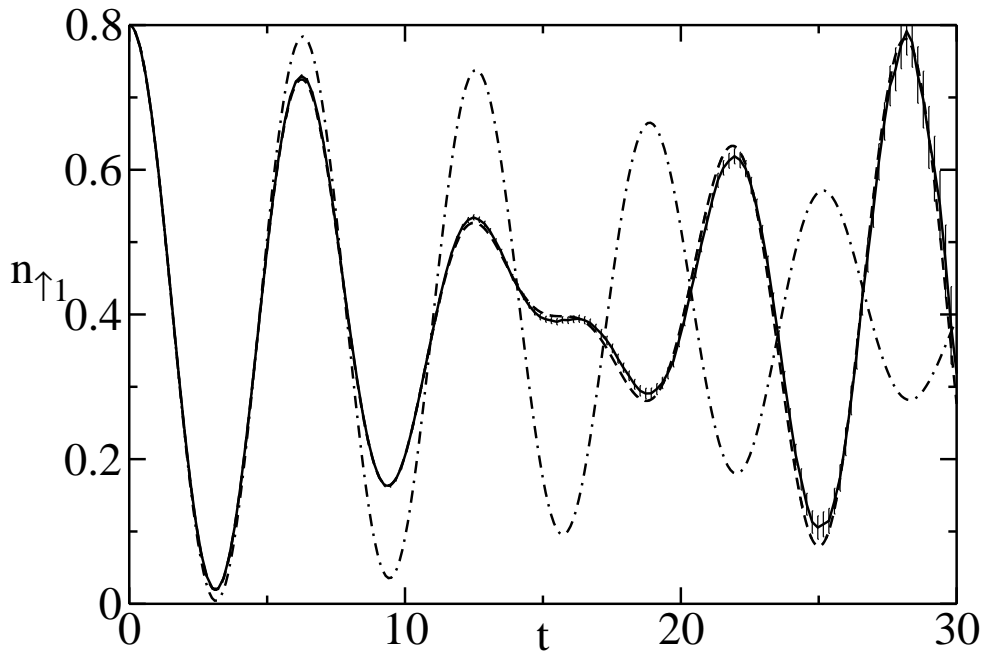,width=11cm,angle=-90}
\caption[]{Mean value of the population in the state $1\uparrow$ as a function of time. The solid and
dashed lines are evaluated using the schemes of \S\ref{subsubsec:optimal} and \S\ref{subsubsec:low_mem}, respectively. 
The number of realizations is $10^5$ and $V=0.2$.
The dashed-dotted line is the BCS mean-field prediction. The widths of the error bars are the standard deviations.}
\label{fig1}
\end{figure}

\begin{figure}
\epsfig{figure=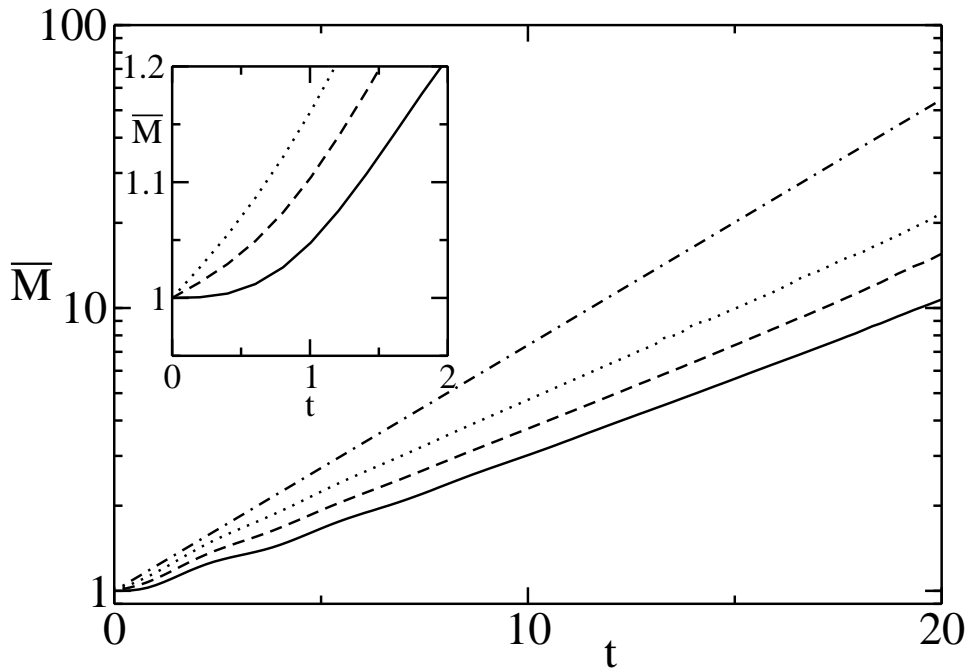,width=11cm,angle=-90}
\caption[]{Average of $M=\langle\Omega,\gamma|\Omega,\gamma\rangle$
over the stochastic realisations, as a function of time, evaluated using the scheme of \S\ref{subsubsec:optimal} (solid line),
of \S\ref{subsec:sol} (dashed line) and of \S\ref{subsubsec:low_mem} (dotted line).
The dashed-dotted 
line is the upper bound of Eq.~(\ref{growth3}) for the first two schemes. At the initial
time the growth of $\overline M$ is zero for the first scheme (see inset).}
\label{fig2}
\end{figure}

At the initial time, we choose a BCS state with the elements of 
$\gamma$ equal to zero, apart from 
$\gamma_{1\uparrow,1\downarrow}=-\gamma_{1\downarrow,1\uparrow}\equiv\gamma_0=2$. The state is a superposition of the 
vacuum and the state with two atoms in the site $1$.
The direct numerical solution of the dynamics is obtained writing the Hamiltonian in the basis of the
Fock states of the operators $\hat c_{i,s}$ and $\hat c_{i,s}^\dagger$. The integration is simplified 
by the fact that the interaction cannot flip the spin and the amplitude of some states remains zero.
In Fig.~\ref{fig1} we report the mean value of the population in the state $\uparrow 1$ as a function
of time. We have set $V_0=0.2$. The dashed-dotted line is evaluated with the mean-field equations (as given in Appendix~\ref{appen:mf}),
the dashed line 
is the direct numerical solution and the solid line is the stochastic solution. The widths of the error bars are the standard
deviations. We have used $10^5$ realizations with the scheme of \S\ref{subsubsec:optimal}. In the mean-field approximation, 
the evolution has a damped oscillation 
with a revival for $t>30$. The collapse and revival of the oscillations of the exact solution occur with a shorter 
time scale. The stochastic approach is able to display very well this behavior. In Fig.~\ref{fig2}
we report $\overline{\langle\Omega,\gamma|\Omega,\gamma\rangle}$ as a function of time. The solid 
dashed and dotted lines are evaluated using the schemes of \S\ref{subsubsec:optimal}  
\S\ref{subsec:sol} and \S\ref{subsubsec:low_mem}, respectively.
The dashed-dotted line is the upper bound of Eq.~(\ref{growth3}) for the first two schemes.
As expected, the optimized scheme has a smaller spreading.

Note that the growth rate of $\overline M$ is zero at the initial time for the scheme of \S\ref{subsubsec:optimal} (see
inset of Fig.~\ref{fig2}), because 
of the presence of the last term in Eq.~(\ref{growth3}), which cancels the other contributions
in the initial state considered here of particles localized on a site. 
The spreading of the trajectories grows exponentially and
it increases for larger interparticle interactions. We have done similar calculations for a stronger
interaction, e.g.\ for $V=2$; the stochastic method agrees with the direct numerical solution,
with a higher growth rate of the statistical error for increasing $V$, as expected.

\section{conclusions}
In this article we have shown that the state evolution of a fermionic gas with binary interactions can
be obtained in an exact way as the average of stochastic trajectories of BCS states. We have derived the general
Ito stochastic equations which give the exact evolution of the system and we have found a condition
on some parameters of these equations
to reduce the statistical spreading of the trajectories in the Hilbert space. The upper bound
that we have found on the spreading for a particular scheme is similar to the one obtained for the 
Hartree-Fock ansatz in \cite{juillet}, with a smaller value.
We have illustrated the method on a two-site model and we have shown that the quantum
effects, which cannot be obtained with a mean-field approximation, are displayed by the results
of the stochastic approach.

\acknowledgments
We acknowledge useful discussions with Jean Dalibard and Iacopo Carusotto at an early stage of
this work. Laboratoire Kastler Brossel is a research unit of Ecole normale sup\'erieure
and of Universit\'e Pierre et Marie Curie, associated to CNRS.
\appendix

\section{Some properties of BCS states}
\label{sect1}
First we prove that the BCS states form a complete family for the states
with an even number of atoms. The set of states with a definite number of atoms 
in each mode constitutes an orthonormal basis of the Hilbert space.
It is sufficient to show that each element of this set is equal to a 
superposition of BCS states.
An element has the following form
\be\label{ele_basis}
|\{k_n,l_n\}\rangle\equiv
\big(\prod_{n=1}^{N_p}\hat c_{k_n}^\dagger\hat c_{l_n}^\dagger\big)|0\rangle,
\ee
where we have grouped the atoms in pairs. The $n$-th pair has the atoms in
the $k_n$ and $l_n$ modes. $N_p$ is the number of pairs. 
$k_n\neq k_m\neq l_n\neq l_m$ when $n\neq m$, whereas $k_n\neq l_m$ for every
$n$ and $m$.
It is easy to prove that
\be
|\{k_n,l_n\}\rangle=K\int_0^{2\pi} d\phi_1 \int_0^{2\pi}
\ d\phi_2...\int_0^{2\pi}d\phi_{N_p}e^{-{\mathrm i}\sum_{n=1}^{N_p}[\phi_n+
\exp({\mathrm i}\phi_n)\hat c_{k_n}^\dagger \hat c_{l_n}^\dagger]}|0\rangle,
\ee
where $K$ is a normalization constant. Thus, the BCS states form a complete
family. Actually, it is over-complete.

It is possible to demonstrate that~(see Section 2.2 of Ref.~\cite{Blaizot}) \cite{deter}
\be\label{norma}
\tilde M\equiv\langle\bar\Omega,\bar\gamma|\Omega,\gamma\rangle=
\bar\Omega^*\Omega\det[\openone+\bar\gamma^\dagger\gamma]^{1/2}
\ee
and
\bey\label{pro1}
\frac{\partial}{\partial\gamma_{kl}}|\Omega,\gamma\rangle=\hat c_k^\dagger
\hat c_l^\dagger|\Omega,\gamma\rangle, \\
\label{pro2}
\hat c_k|\Omega,\gamma\rangle=\sum_l\gamma_{kl}\hat c_l^\dagger|\Omega,
\gamma\rangle.
\eey
 From Eq.~(\ref{norma}) we have 
\be\label{deriv}
\frac{\partial}{\partial\gamma_{ij}}\tilde M
=-[\bar\gamma^\dagger(\openone+\gamma\bar\gamma^\dagger)^{-1}]_{ij}
\tilde M.
\ee
Using Eqs.~(\ref{pro1}-\ref{deriv}) we find that
\bey
\frac{\langle\bar\Omega,\bar\gamma|\hat c_l\hat c_k^\dagger
|\Omega,\gamma\rangle}{\langle\bar\Omega,\bar\gamma|\Omega,\gamma\rangle} &=&
[(\openone+\bar\gamma^\dagger\gamma)^{-1}]_{kl} \\
\label{cpc}
\frac{\langle\bar\Omega,\bar\gamma|\hat c_k^\dagger\hat c_l
|\Omega,\gamma\rangle}{\langle\bar\Omega,\bar\gamma|\Omega,\gamma\rangle}&=&
[\bar\gamma(\openone+\bar\gamma^\dagger\gamma)^{-1}\gamma^\dagger
]_{lk}
\\
\label{cc}
\frac{\langle\bar\Omega,\bar\gamma|\hat c_k\hat c_l|
\Omega,\gamma\rangle}{\langle\bar\Omega,\bar\gamma|\Omega,\gamma\rangle}&=&
-[\gamma(\openone+\bar\gamma^\dagger\gamma)^{-1}]_{kl} \\
\frac{\langle\bar\Omega,\bar\gamma|\hat c_l^\dagger
\hat c_k^\dagger|\Omega,\gamma\rangle}{\langle\bar\Omega,\bar\gamma|
\Omega,\gamma\rangle}&=&
[\bar\gamma^\dagger(\openone+\gamma\bar\gamma^\dagger)^{-1}]_{kl}.
\eey
We note that Equation~(\ref{cpc}) can be written in various forms using
the matrix identities
\be\label{matrid}
\bar\gamma^\dagger\gamma(\openone+\bar\gamma^\dagger\gamma)^{-1}=
\bar\gamma^\dagger(\openone+\gamma\bar\gamma^\dagger)^{-1}\gamma=
[\gamma(\openone+\bar\gamma^\dagger\gamma)^{-1}\bar\gamma^\dagger]^T
\ee
where $A^T$ is the transpose of matrix $A$.

\section{Mean-field equations}
\label{appen:mf}

Using the results of section 9.9b of \cite{Blaizot}, 
we obtain the following equations of motion for
$\gamma$ in the mean-field approximation:
\be
\dot\gamma_{ij}=-{\mathrm i}\sum_k(h_{ik}\gamma_{kj}+h_{jk}\gamma_{ik})
+\left[-{\mathrm i}\sum_kV_{ik}\langle\hat{c}_k^\dagger\hat{c}_k\rangle\gamma_{ij}+{\mathrm i}
\sum_kV_{ik}\langle\hat{c}_k^\dagger\hat{c}_i\rangle\gamma_{kj}-(i\leftrightarrow j)\right]
+{\mathrm i}\sum_{kl}V_{kl}\langle\hat{c}_k\hat{c}_l\rangle^*\gamma_{ki}\gamma_{jl}+{\mathrm i}
V_{ij}\langle\hat{c}_i\hat{c}_j\rangle.
\ee


\begin{thebibliography}{}
\bibitem{why_not_bad}
One may be worried by the fact that the stochastic ansatz does not have a well
defined number of particles. E.g.\ for bosons it was shown that using a Glauber coherent
state ansatz (which is the bosonic analog of the BCS state) leads to divergences
in the statistical error \cite{caru1}. However, such a pathology cannot occur
for fermions with a finite number of modes: the total number of fermions is bounded from above by the number
of modes in the system. In a BCS ansatz, one has even that the variance of the
number of particles is less than twice the mean number of particles.
Furthermore, if the initial BCS state of the fermions corresponds to the limit
of pairs localized on individual lattice sites, one shall find in the optimized scheme
of \S\ref{subsubsec:optimal}  that the initial growth
rate of the statistical error is zero. This is not the case for the stochastic coherent 
scheme for bosons for an initial coherent state localized on a single lattice site.
\bibitem{how_large}
In \cite{juillet} one has to write the interaction term
as $\sum_s \hbar\omega_s O_s^2/4$, where $O_s$ are Hermitian one-body
operators.
A natural choice is to take $s=r$ and $O_r=\hat{c}_{\uparrow r}^\dagger
\hat{c}_{\uparrow r}-\hat{c}_{\downarrow r}^\dagger \hat{c}_{\downarrow r}$,
and $\hbar\omega_r=-2 V_0$. When expanding the square $O_r^2$ one indeed
uses the fact (for fermions) 
that the square of an occupation number is equal to the 
occupation number, which allows to reincorporate the undesired terms
into the one-body part of the Hamiltonian. From Eq.(19) of
\cite{juillet} one then gets $M(t)\leq \exp(t |V_0| m_s)$,
so that the rate inside the exponent is two times larger than the
one for the stochastic BCS ansatz.
\bibitem{suff} Note that imposing such a correlation
function on the $f_i$'s is sufficient but not necessary; this is exemplified
by the fact that the two other schemes of this section, though of
the form Eq.~(\ref{ansate}), lead to $\Delta f_i$'s with different
correlation functions.
\bibitem{aide} To get the upper bound, we used the fact that $n_k (1- n_k)\leq 1/4$
where $n_k = \langle \hat{c}_k^\dagger \hat{c}_k\rangle$.
\bibitem{deter} The identity Eq.~(\ref{norma}) determines $\tilde{M}$ up to a sign.
For the numerical example given in this paper, we have checked formally that
the determinant in Eq.~(\ref{norma}) can be written as a square of a polynomial 
in the $\gamma_{ij}$'s and the $\bar{\gamma}_{ij}$'s, which allows to choose
the correct determination by continuation from $\bar{\gamma}=0$.
\end{thebibliography}
\end{document}